\documentclass[apj,numberedappendix]{emulateapj}
\usepackage{apjfonts,graphicx,url,amssymb,amsmath,longtable,rotating,color,units,wasysym,subfigure,epsfig,listings,bm}
\usepackage{graphicx,amssymb,amsmath,epsfig}

\slugcomment{Submitted for publication in The Astrophysical Journal Letters}
\shorttitle{Transparent Helium} 
\shortauthors{A. L. Piro and V. S. Morozova}

\newcommand{\be}{\begin{eqnarray}}
\newcommand{\ee}{\end{eqnarray}}
\newcommand{\lp}{\left(}
\newcommand{\rp}{\right)}

\begin{document}


\title{Transparent Helium in Stripped Envelope Supernovae}

\author{Anthony L. Piro and Viktoriya S. Morozova}

\affil{Theoretical Astrophysics, California Institute of Technology, 1200 E California Blvd., M/C 350-17, Pasadena, CA 91125, USA; piro@caltech.edu}


\begin{abstract}
Using simple arguments based on photometric light curves and velocity evolution, we propose that some stripped envelope supernovae (SNe) show signs that a significant fraction of their helium is effectively transparent. The main pieces of evidence are the relatively low velocities with little velocity evolution, as are expected deep inside an exploding star, along with temperatures that are too low to ionize helium. This means that the helium should not contribute to the shaping of the main SN light curve, and thus the total helium mass may be difficult to measure from simple light curve modeling. Conversely, such modeling may be more useful for constraining the mass of the carbon/oxygen core of the SN progenitor. Other stripped envelope SNe show higher velocities and larger velocity gradients, which require an additional opacity source (perhaps the mixing of heavier elements or radioactive nickel) to prevent the helium from being transparent. We discuss ways in which similar analysis can provide insights into the differences and similarities between SNe Ib and Ic, which will lead to a better understanding of their respective formation mechanisms.
\end{abstract}



\section{Introduction}

One of the fundamental problems in stellar astrophysics is connecting supernovae (SNe) with their massive stellar progenitors. There is strong evidence that Type II-P SNe are from the core-collapse of red supergiants, both via direct identification with pre-explosion imaging \citep{Smartt09} and modeling of their light curves \citep{Falk77,Eastman94,Kasen09,Dessart10,Bersten11,Dessart11}. In contrast, for the mass stripped SNe (Type Ib, Ic, and IIb), direct identification has been difficult with the exception of a few cases of yellow supergiants associated with SNe IIb \citep{Maund11,Bersten12,VanDyk13,VanDyk14}. Historically, there has been debate whether their mass stripping comes from the winds of effectively isolated stars or if it is due to binary interactions. However more recently, both light curve modeling \citep{Ensman88,Dessart11b,Benvenuto13,Bersten14,Fremling14}, which favors relatively low ejecta masses, and the high rate of SNe Ib and Ic \citep{Smith11} argue that the binary origin explains the majority of these events \citep[also see][]{Smith14}.

A further complication in trying to understand the origin of mass stripped SNe is identifying the mechanism that determines whether an SN is of Type Ib or Ic. Spectroscopically, this difference is simply attributed to a lack of observed helium, but as highlighted by \citet{Dessart12}, this does not necessarily mean that SNe Ic are intrinsically helium poor \citep[although see][]{Hachinger12}. Non-thermal excitation and ionization are key for the production of He I lines \citep{Lucy91}. Thus an SN progenitor with helium-rich surface layers could in principle look like an SN Ib or Ic depending on the amount of mixing of $^{56}$Ni.
Given these complications, it would be useful to have simple rules of thumb to determine what these SNe are telling us about the presence or not of helium and understand how it impacts other inferences about the progenitor. This is the motivation of the present work.

A further motivation is the steadily growing sample of stripped envelope SNe \citep{Drout11,Modjaz14} that will continue to be discovered by current and future surveys, such as PTF \citep{Rau09}, Pan-STARRS \citep{Kaiser02}, CSP \citep{CSP}, LCOGT \citep{LCOGT}, ZTF \citep{Law09}, ASAS-SN \citep{Shappee13b}, and LSST \citep{LSST}. This allows events to be studied in aggregate to search for interesting trends in the ejecta masses, energetics, $^{56}$Ni, and a range of other properties \citep{Lyman14}. Such work is well-suited for simple modeling to control the parameter space, but it is only useful if the limitations of such modeling are properly understood.

In Section \ref{sec:color velocity}, we demonstrate that the He I lines in three example SNe IIb/Ib are a useful tracer for their photospheres. This shows that at least for the two SNe IIb we consider, we are likely seeing deep into the SN ejecta. We make comparisons with hydrodynamic models of exploding stars to strengthen this case. Coupled with the temperature evolution and the opacity of the helium-rich surface layers presented in Section \ref{sec:temperature}, we argue that a non-negligible fraction of helium is recombined in these SNe IIb. We conclude in Section \ref{sec:conclusions} with a summary of our conclusions and a discussion of future work, particularly for comparing Type Ib and Ic SNe.

\section{Color Velocity Evolution}
\label{sec:color velocity}

As a shock passes through an exploding star, it accelerates and unbinds material. The velocity to which ejecta is accelerated can vary greatly throughout the star and is sensitive to the stellar density profile as described in \citet{Matzner99}. Roughly speaking, the velocity decreases as the shock moves out and sweeps up more mass, but it can also accelerate when the density decreases rapidly, especially at the star's surface. Therefore, high velocities and large velocity gradients indicate that observations are probing material with a large density gradient, likely near the surface of the star.

\begin{figure}
\epsscale{1.2}
\plotone{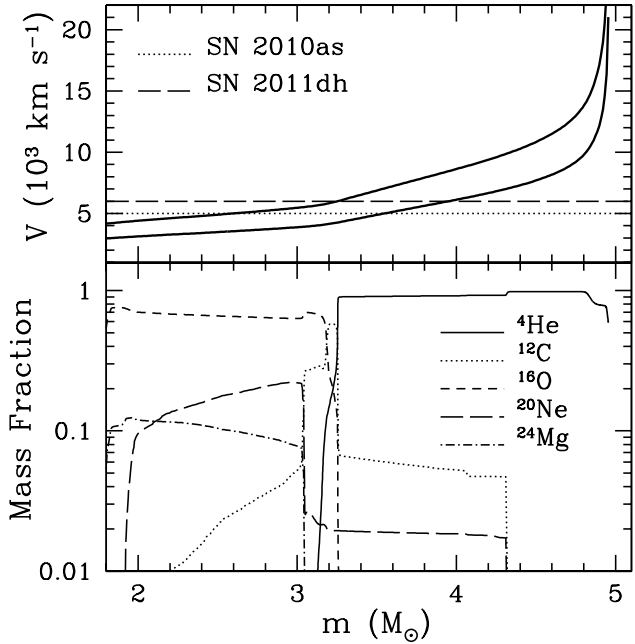}
\caption{Profile of an exploding star as a function of mass coordinate. The model is a $15\,M_\odot$ zero-age main-sequence star with its hydrogen envelope removed to simulate binary mass loss. The upper panel shows terminal velocity profiles of $10^{51}$ and $2\times10^{51}\, {\rm erg}$ explosions (solid lines, bottom and top, respectively). The horizontal lines show typical He I velocities for SNe 2010as and 2011dh to help guide the eye. The bottom panel shows the composition of the most abundant elements at the moment of core-collapse.}
\epsscale{1.0}
\label{fig:profile}
 \end{figure}

These features can be seen in Figure \ref{fig:profile}, which shows in the upper panel the terminal velocity profile, for a range of different explosion energies, in a star that has a mass of $\sim5\,M_\odot$ at core collapse. This star was generated from a $15\,M_\odot$ zero-age main-sequence star using the 1D stellar evolution code \texttt{MESA} \citep{Paxton13}. Using the overshooting and mixing parameters recommended by \citet{Sukhbold14}, the star is evolved until a large entropy jump between the core and envelope was established. The convective envelope is removed to mimic mass loss during a common envelope phase, and then the star continues to evolve up to core collapse. A shock is initiated by heating the star at a mass coordinate of $m=1.5\,M_\odot$, and the subsequent hydrodynamics evolution is followed using our 1D Lagrangian supernova explosion code (\texttt{SNEC}, Morozova et al. in preparation, which follows the numerical hydrodynamic scheme of \citealp{Mezzacappa93}). The main features to note are the great acceleration of the shock near the surface and the more modest acceleration at the boundary between the carbon/oxygen layers and the helium-rich surface layers. In the bottom panel of Figure \ref{fig:profile}, we plot the composition of some of the most abundant elements in the star prior to explosion.

\begin{figure}
\epsscale{1.2}
\plotone{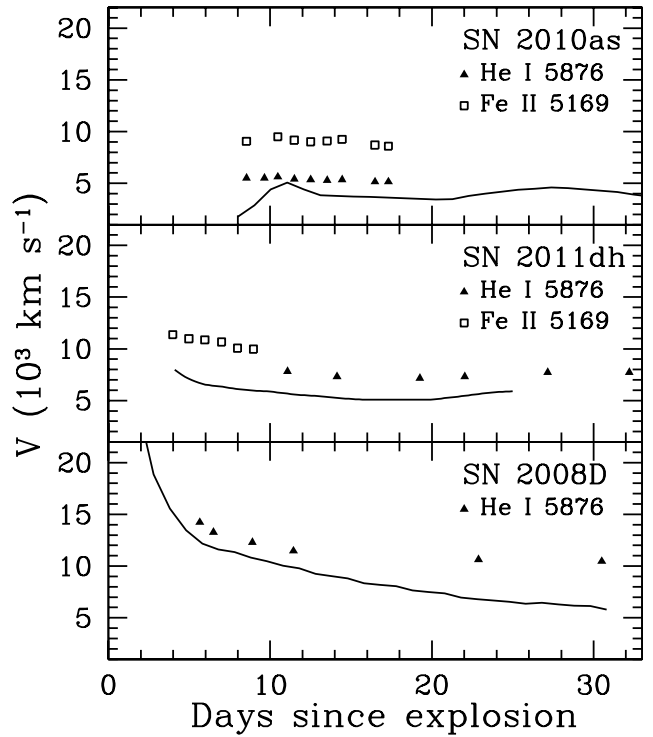}
\caption{The solid line in each case marks $V_c$ in the SN that was found using Equation (\ref{eq:v}) with $\tau_c=1$. The upper two panels are Type IIb SNe that have been noted for their especially low velocity helium, while the bottom panel is a SN Ib with a more typical velocity evolution. Nevertheless, in all three cases, $V_c$ is similar but below the helium velocity.}
\epsscale{1.0}
\label{fig:velocities}
\end{figure}

Recently, \citet{Folatelli14} pointed out that some stripped envelope SNe appear to have anomalously low He I velocities in the range of $4,000$ to $8,000\,{\rm km\,s^{-1}}$ with one of the prime examples being SN 2010as (see Figure \ref{fig:velocities}). This is lower than the typical velocities of $10,000$ to $15,000\,{\rm km\,s^{-1}}$ (see the surface layers in Figure \ref{fig:profile}) typically associated with the photosphere. In addition, the evolution with time is much flatter, which is again not expected given the high velocity gradient in the surface layers.

To better understand what is implied by these seemingly low velocities, it is helpful to consider the velocity of material at the color depth (also known as the thermalization depth), henceforth referred to as the color velocity $V_c$. For a black body with temperature $T_{\rm BB}$ and color radius $r_c$, the bolometric luminosity is $L=4\pi r_c^2\sigma_{\rm SB}T_{\rm BB}^4/\tau_c$, where $\sigma_{\rm SB}$ is the Stefan-Boltzmann constant and $\tau_c$ is the optical depth at the color depth. Substituting $r_c=V_ct$,
\be
	V_c = \frac{1}{t}\lp \frac{L\tau_c}{4\pi \sigma_{\rm SB}T_{\rm BB}^{4}}\rp^{1/2}.
	\label{eq:v}
\ee
The color radius $r_c$ is roughly where an incoming photon would experience at least one absorption, rather than just a scattering, and thus where the observed black body temperature is determined
\citep[see][]{Nakar10,Piro13}. The color radius is in general slightly deeper than the actual photosphere. Nevertheless, $V_c$ should provide a useful diagnostic for roughly tracking the photospheric velocity.

Equation (\ref{eq:v}) with $\tau_c=1$ is used to analyze three different events, the Type IIb SNe 2010as \citep{Folatelli14} and 2011dh \citep{Marion14}, and the Type Ib SN 2008D \citep{Modjaz09}. Setting $\tau_c=1$ gives a lower limit on $V_c$, since $\tau_c>1$. Although SNe IIb have a thin hydrogen layer at the surface \citep{Woosley94,Bersten12,Nakar14}, this is sufficiently low mass that it will be optically thin at the times we consider and not affect our arguments. In Figure \ref{fig:velocities}, we also plot the measured He I 5876 absorption line velocities and in two cases the Fe II 5169 velocities. In all three SNe, $V_c$ is similar but less than the He I velocity. This argues that indeed the helium is tracking the photosphere, and that $\tau_c$ cannot be too far from unity. We conclude that even when the velocity is low there is nothing intrinsically anomalous about the He I velocities. Rather they are simply consistent with the velocity of the ejecta as inferred from the luminosity and temperature.

As a comparison, in the upper panel of Figure \ref{fig:profile} we plot horizontal lines roughly at the He I velocities for SNe 2010as and 2011dh. In both cases, these velocities are similar to what is expected near the boundary between the carbon/oxygen inner layers and the helium-rich surface layers. Furthermore, one can see that the velocity profile near these regions is much shallower than the surface velocity, again consistent with the observed He I features. Although this comparison cannot provide a quantitative result, at least qualitatively it appears that for SNe 2010as and 2011dh we are looking deep into the ejecta. In contrast, SN 2008D shows  higher velocities with a larger velocity gradient consistent with the outer portions of the star, and thus we are not looking as deep into the ejecta.

\section{Temperature and Opacity}
\label{sec:temperature}

\begin{figure}
\epsscale{1.2}
\plotone{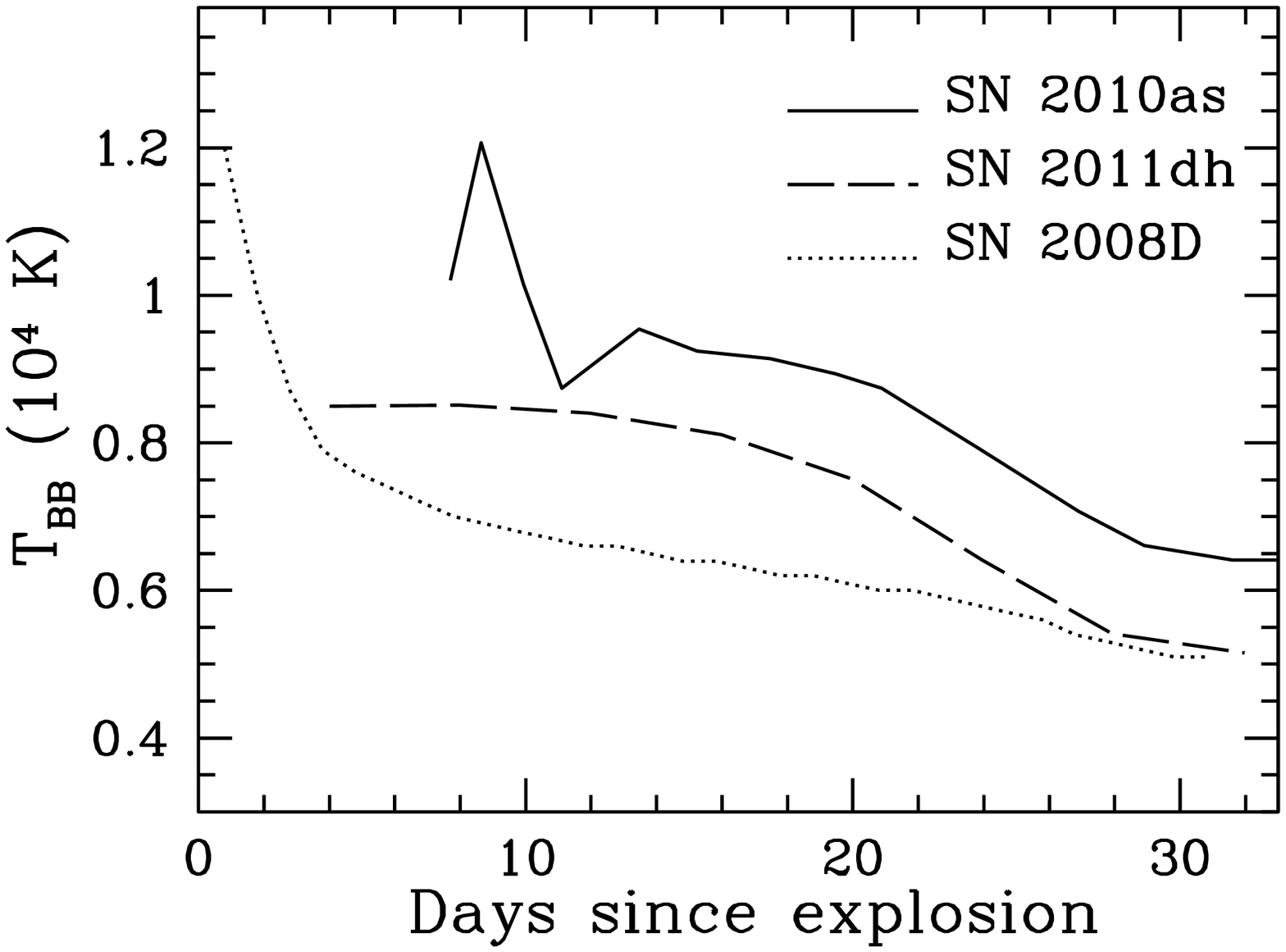}
\caption{Black body temperature as a function of time for each of the three SNe, using the data from \citet{Folatelli14}, \citet{Marion14}, and \citet{Modjaz09}. }
\epsscale{1.0}
\label{fig:temperature}
\end{figure}

\begin{figure}
\epsscale{1.2}
\plotone{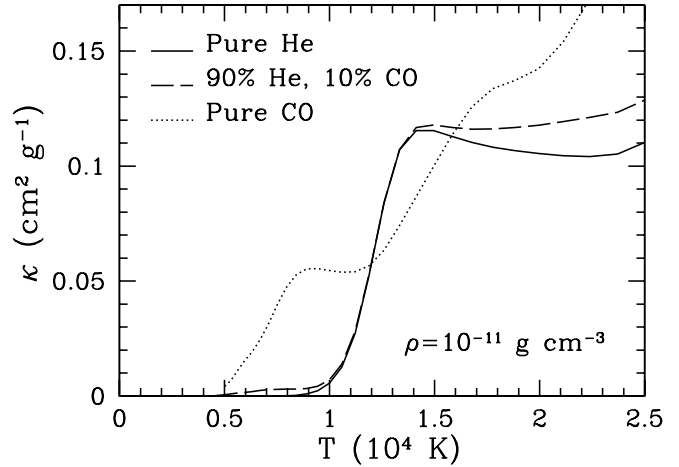}
\caption{The Rosseland mean opacity as a function of temperature and at a representative density of $10^{-11}\,{\rm g\,cm^{-3}}$ ($\kappa$ is weakly dependent on density). We plot a composition of pure helium (solid line), helium with a 10\% mass fraction carbon/oxygen (dashed line), and pure carbon/oxygen (dotted line).}
\epsscale{1.0}
\label{fig:kappa}
 \end{figure}

To understand why the He I and $V_c$ have relatively small velocities, it is helpful to consider the actual black body temperatures that are being observed. In Figure \ref{fig:temperature}, we plot the temperature evolution for each of the three SNe. In comparison, in Figure \ref{fig:kappa}, we plot the opacity as a function of temperature for helium-rich material \citep[][and references therein]{Badnell05}, as is expected in the outer layers of these stripped envelope progenitors. Above a temperature of $\approx1.2\times10^4\,{\rm K}$, the opacity is $\approx0.1\,{\rm cm^2\,g^{-1}}$, consistent with electron scattering from material with one electron per four nucleons (i.e., singly ionized helium). Below this temperature, the opacity is almost zero and the material is effectively transparent.  We also consider a mixture with 10\% carbon and oxygen. This particular plot uses a density of $10^{-11}\,{\rm g\,cm^{-3}}$, but the threshold at which this happens is not strongly dependent on density. Only once the material is largely carbon/oxygen does the opacity increase much for low $T$.

Combining Figures \ref{fig:temperature} and  \ref{fig:kappa} with the low and flat He I velocities, we conclude that a large fraction of helium in SNe 2010as and 2011dh is transparent. To be clear, it is true that helium absorption lines are seen in all of these SNe, but this is only at a few specific wavelengths. What we are arguing is that across the majority of the spectrum, the helium-rich material is not drastically impacting the time it takes photons to diffuse out of the expanding ejecta. One potential complication is if there are opacity sources not taken into account in Figure \ref{fig:kappa} that increase the opacity. A standard practice for light curve modeling codes is to invoke an opacity floor \citep[][and references therein]{Bersten11}. This replicates physics such as bound-free and bound-bound absorptions, and non-thermal excitation or ionization of electrons by Compton scattering of $\gamma$-rays (although \citealp{Kleiser14} show that at least the bound-bound opacity for pure-helium is negligible). Nevertheless, the close match between the He I line velocities and $V_c$ in Figure \ref{fig:velocities} argues that this opacity floor cannot be too large and that a non-negligible fraction of the helium is recombined and not providing a large opacity. Although SN 2008D has even colder temperatures, it does not have the same low velocities or flat velocity evolution. This SN therefore requires an additional opacity source within its helium-rich surface layers, such as the mixing of metals or $^{56}$Ni, the latter of which could assist in helium ionization. Calculations using an opacity floor should therefore be more appropriate for events with velocities like SN 2008D.

The inference of transparent material has implications for studies that use simple models to constrain the properties of the SNe from their photometric light curves. The basic idea of these works is to relate the rise time of the SN to the ejecta mass, $t_{\rm rise} \propto (\kappa M_{\rm ej}/v_{\rm ej}c)^{1/2}$, where $M_{\rm ej}$ and $v_{\rm ej}$ are the typical mass and velocity of the ejecta, respectively \citep{Arnett82}. This should work as long as the opacity $\kappa$ is representative of the majority of the material, but not if $\kappa$ varies by a large amount as happens if some material is transparent. Recently \citet{Lyman14} used similar semi-analytic methods to study the ejecta masses in 38 Type IIb, Ib, Ic, and Ic-BL SNe. One particularly striking conclusion was that the majority of these stripped envelope SNe have rather similar ejecta masses in the range of $\sim1-5\,M_\odot$. If the typical understanding of the difference between SNe Ib and Ic were true, namely that SNe Ic have additional mass loss to remain their helium envelopes, one would instead expect on average larger ejecta masses from SNe Ib. Our discussion here demonstrates that indeed the SNe Ib could have more ejecta overall, but the mass of the helium does not impact the light curve width because it is transparent. What these measurements are able to constrain with certainty is the mass of the progenitor's carbon/oxygen core. In the future, consideration of the velocity evolution of each SN should be taken into account when assessing each inferred ejecta mass.

\section{Conclusion and Discussion}
\label{sec:conclusions}

We have inferred the time-dependent color velocity $V_c$ for three different SNe and compared it with hydrodynamic models of mass stripped star explosions. We then argued that the combination of (1) black body temperatures too low for helium ionization, (2) relatively low He I velocities with a flat evolution in time, and (3) a $V_c$ evolution that indicates the He I velocities are representative of the photosphere, together suggest that a non-negligible amount of helium is effectively transparent in SNe 2010as and 2011dh. This could result in a solar mass or more being missed by simple models attempting to infer the ejecta mass. {\em Even in detailed numerical studies of these events, constraining the total ejecta mass may be difficult.} This is because even if a given model is shown to fit, a similarly good fit may be possible with an additional amount of helium (which may require corresponding adjustments to the explosion energy). Future numerical modeling needs to quantify just how much helium can actually be hidden.

In contrast, SN 2008D has higher velocities and a larger velocity gradient consistent with material near the surface of a star, even though its $T_{\rm BB}$ is lower than the other two SNe. This indicates an additional opacity source is required to prevent the helium from being more transparent. Our results will hopefully motivate similar analysis of $V_c$ in future SN studies. This was recently done for the SN Ib iPTF13bvn \citep{Fremling14}, and again $T_{\rm BB}<10^4\,{\rm K}$ and $V_c$ is similar to the He I. This work constrained the ejecta mass to be $\approx2\,M_\odot$ \citep[also see][]{Bersten14}, and thus would rule out a Wolf-Rayet progenitor \citep[as suggested by pre-explosion photometry,][]{Cao13}. Although transparent helium may not be enough to reconcile this ejecta mass difference with a Wolf-Rayet star, it could easily be a significant correction.

To maximize the usefulness of such studies, it will be important to have measurements of $L$, $T_{\rm BB}$, and $t$, so that Equation (\ref{eq:v}) can be applied. This puts emphasis on covering all wavelengths so that the bolometric luminosity and black body temperature can be accurately inferred \citep[see the discussion in][]{Marion14}. Infrared can be helpful for correctly measuring the He I velocities since at optical wavelengths He I absorption features can potentially be confused with other elements (although in any case the lowest securely observed He I velocity should be used to get as close to the photosphere as possible). Getting an accurate $t$ requires that the SN be detected as early as possible, since in cases where a first peak in the light curve is not seen it can be difficult to know exactly when the explosion happened \citep{Piro13}.

Comparing and contrasting this type of analysis between SNe Ib and Ic may be especially instructive for understanding their respective origins. The SN shock will always accelerate material to high velocities near the surface of the star, as shown in Figure \ref{fig:profile}.  {\em If SNe Ic are truly devoid of helium, then these large velocities will occur in higher opacity carbon/oxygen material rather than lower opacity helium material.} This should give rise to $V_c\approx10,000-15,000\,{\rm km\,s^{-1}}$ with a larger velocity gradient as is seen for SN 2008D.  If SNe Ic always show such a $V_c$ evolution that is consistent with no helium present, this would be strong evidence that SNe Ic experience more mass stripping than SNe Ib.
If instead there is evidence for transparent helium in some SNe Ic (i.e., they show $V_c$ evolution similar to SNe 2010as and 2011dh), this would argue that there is transparent helium material present. In this case, the relative deposition of helium versus $^{56}$Ni may be an important factor for deciding the classification, as described by \citet{Dessart12}, so that issues like mixing, turbulence, rotation, and asymmetries need to be more fully considered for generating a Type Ib or Ic. No matter the solution, determining between these scenarios would be an important step forward for our understanding of stripped envelope SNe.

\acknowledgements
We thank Drew Clausen for generating the $15\,M_\odot$ model used for Figure \ref{fig:profile}, Melina Bersten and Mattias Ergon for helpful discussions, and Dan Kasen, Ehud Nakar, Christian Ott, and Nathan Smith for feedback on previous drafts. We also thank the Carnegie Supernova Project, and in particular Mark Phillips, for generously hosting ALP at the CSP II Team Meeting at St. George Island, Florida where this work was inspired. This work was supported through NSF grants AST-1205732, PHY-1068881, PHY-1151197, PHY-1404569, and the Sherman Fairchild Foundation.


\bibliographystyle{apj}
\bibliography{ms}
\end{document}